\documentclass[notitlepage,a4paper,aps,prd,onecolumn,superscriptaddress,nofootinbib,groupedaddress]{revtex4}
\usepackage{enumerate}
\usepackage{amsmath}
\usepackage{amsfonts}
\usepackage{amssymb}
\usepackage[utf8]{inputenc}
\usepackage[T1]{fontenc}
\usepackage{mathtools}
\usepackage{wasysym}

\newcounter{subeq}

\DeclareMathAlphabet{\mathds}{U}{BOONDOX-ds}{m}{n}
\usepackage[dvipsnames]{xcolor}

\usepackage[colorlinks=true]{hyperref}
\usepackage{amsthm}

\theoremstyle{definition}

\theoremstyle{plain}

\newcommand{\dd}{\mathrm{d}}

\allowdisplaybreaks

\begin{document}
\title{Photon sphere and perihelion shift in weak $f(T)$ gravity}

\author{Sebastian Bahamonde}
\email{sbahamonde@ut.ee}
\affiliation{Laboratory of Theoretical Physics, Institute of Physics, University of Tartu, W. Ostwaldi 1, 50411 Tartu, Estonia.}
\affiliation{Department of Mathematics, University College London, Gower Street, London, WC1E 6BT, United Kingdom}

\author{Kai Flathmann}
\email{kai.flathmann@uni-oldenburg.de}
\affiliation{Department of Physics, University of Oldenburg, 26111 Oldenburg, Germany.}

\author{Christian Pfeifer}
\email{christian.pfeifer@ut.ee}
\affiliation{Laboratory of Theoretical Physics, Institute of Physics, University of Tartu, W. Ostwaldi 1, 50411 Tartu, Estonia.}

\begin{abstract}
Among modified theories of gravity, the teleparallel $f(T)$ gravity is an intensively discussed model in the literature. The best way to investigate its viability is to derive observable predictions which yield evidence or constraints for the model, when compared with actual observations. In this paper we derive the photon sphere and the perihelion shift for weak $f(T)$ perturbations of general relativity. We consistently calculate first order teleparallel perturbations of Schwarzschild and Minkowski spacetime geometry, with which we improve and extend existing results in the literature.
\end{abstract}

\maketitle

\section{Introduction}\label{sec:intro}
With the observation of gravitational waves of merging black holes~\cite{Abbott:2016blz} and the first direct picture of a black hole shadow in the center of the galaxy M87~\cite{Akiyama:2019cqa}, the possibilities to observe the behavior of gravity in the strong field regime has increased enormously. The newly obtained data is the perfect basis to understand the viability range of general relativity and possible modified gravity theories, suggested as its generalization. Here we derive the influence of a teleparallel modification of general relativity on the photon sphere of black holes and on the perihelion shift of elliptic orbits in spherical symmetry. This is a step to investigate the influence of teleparallel gravity on more realistic spinning black holes with axial symmetry.

Teleparallel theories of gravity are formulated in terms of a tetrad of a spacetime metric and a spin connection, instead of in terms of a spacetime metric and its Levi-Civita connection~\cite{Aldrovandi:2013wha}. This structure allows for the construction of a huge variety of theories of gravity beyond general relativity, among them the most famous model, the so-called $f(T)$ gravity~\cite{Cai:2015emx,Ferraro:2006jd}. In this theory, the Lagrangian is given by an arbitrary function $f$ of the torsion scalar $T$, which defines the teleparallel equivalent formulation of general relativity (TEGR). Numerous viability criteria for $f(T)$ gravity have been derived in the context of cosmology~\cite{Bengochea:2008gz,Cai:2011tc,Bamba:2010wb,Dent:2011zz}. However, not much work has been done in spherical and axial symmetry, mostly due to the lack of analytic solutions of the field equations. To solve the $f(T)$ gravity field equations in spherical symmetry in all generality for arbitrary $f$ is a difficult task. The main challenge is to find the tetrad, to which one can consistently associate a vanishing spin-connection. This tetrad does not only have to satisfy the symmetric part and anti-symmetric part (the spin-connection part) of the field equations~\cite{Tamanini:2012hg}, but also must yield a torsion scalar which vanishes for the Minkowski spacetime limit. A further subtlety is that some solutions of the $f(T)$ field equations yield a constant torsion scalar $T$. In this case $f(T)$ gravity is identical to TEGR plus a cosmological constant and nothing new is obtained. The latter feature is for example present in the first study which tried to find spherically symmetric solutions~\cite{Ferraro:2011ks} and also in a later study which used the Noether's symmetry approach to find solutions~\cite{Paliathanasis:2014iva}.

There are only a few publications deriving exact solutions with the correct field equations (see for example~\cite{Daouda:2012nj}). In addition, some regular black hole solutions (perturbatively and exact) have been found correctly in~\cite{Aftergood:2014wla,Boehmer:2019uxv}. When one considers matter, there are some works which have studied the possibility of constructing stars or wormhole solutions in different teleparallel theories of gravity~\cite{Bohmer:2011si,Jamil:2012ti,Bahamonde:2016jqq,Jawad:2015uea,Horvat:2014xwa,Boehmer:2011gw,Ilijic:2018ulf,Pace:2017dpu,Pace:2017aon}. Overall the issue of finding exact spherically symmetric solutions in $f(T)$-gravity is still an open problem.

Instead of looking for full analytical solutions, an alternative way to study the astrophysical effects of modified theories of gravity is to employ perturbation theory. The influence of deviations from TEGR can be investigated, by setting $f(T) = T + (1/2)\epsilon \alpha T^p$. This model contains TEGR (and GR) in the limit the perturbation parameter $\epsilon$ or the coupling constant $\alpha$ goes to zero. It is assumed that the deviation from TEGR is small, ($\epsilon \ll 1$), and hence, only first order terms in $\epsilon$ are relevant in all calculations. In this paper, we consider perturbations around two different background geometries: Minkowski spacetime and Schwarzschild spacetime. 

For the first case we keep the exponent $p >1/2$ and solve the spherically symmetric perturbative $f(T)$ field equations in vacuum. We find that, to first order, the teleparallel perturbation of general relativity has no influence at all. This finding is in conflict with results found earlier in \cite{Ruggiero:2015oka}. A problem with these earlier derivations is that the $f(T)$ field equations presented in \cite{Ruggiero:2015oka}, Eqs.~(8) through (10), do not have Schwarzschild geometry as a solution for $f(T) = T$ and vanishing matter. Moreover the tetrad used there does not yield a torsion scalar which vanishes in the Minkowski spacetime limit. These shortcomings made us redo the calculations, while paying particular attention to the consistency of the perturbation theory.

For the second case, the first order field equations are more involved and cannot be solved for general $p$. To find the astrophysical impact of the parameter $p$, we derive perturbative solutions for $p=2$ to $10$, from which we calculate the circular particle orbits for massless particles and the perihelion shift for nearly circular massive particle orbits. The circular orbits of massless particles define the photon sphere, which is interesting in particular, since it defines the edge of the shadow of the black hole. For $p=2$ the perihelion shift has already been studied in the literature and we recover the results from \cite{DeBenedictis:2016aze}. Comparing our calculation with the previous one demonstrates explicitly that the covariant formulation of teleparallel gravity works well. We employ a vanishing spin connection and a non-diagonal tetrad, while in \cite{DeBenedictis:2016aze} a non-vanishing spin-connection and a diagonal tetrad was used.

The main aim of this paper is to present a careful derivation of the first order $f(T)$ field equations in the single tetrad framework for the models mentioned above and to derive the perturbative solutions around Minkowski and Schwarzschild geometry. Eventually this procedure gives insights about the phenomenological consequences of the teleparallel corrections to general relativity. This work prepares a more general study where we will derive the phenomenological consequences of teleparallel perturbations of Kerr geometry, with and without cosmological constant.

The article is structured as follows: In Sec.~\ref{sec:covtelegrav} we give an overview about the covariant formulation of $f(T)$ gravity and then, we find the corresponding field equations for any spherically symmetric spacetime. Sec.~\ref{sec:weak} is devoted to studying the weak power-law f(T) model for perturbations around Minkowski and Schwarzschild geometries to find the correct metric coefficients which solve the first order field equations. The particle motion phenomenology for the squared power-law $f(T)$ case for the Schwarzschild background is studied in Sec.~\ref{sec:pheno}, deriving the deviation from TEGR (or GR) of the photon sphere and the perihelion shift. We conclude our main results in Sec.~\ref{sec:conclusion}

Throughout the paper we denote $h^{a}{}_{\mu}$ and $h_{a}{}^{\mu}$ for the tetrad and its inverse, respectively, where Latin indices refer to tangent space indices and Greek to spacetime indices. Our signature convention is $(+,-,-,-)$ and we work in units where $G=c=1$.

\section{Covariant formulation of $f(T)$ gravity in spherical symmetry}\label{sec:covtelegrav}
Throughout this paper we employ the covariant formulation of teleparallel gravity \cite{Krssak:2015oua} in the Weitzenb\"ock gauge, also called the pure tetrad formalism. That means we consider a tetrad, its torsion, and a vanishing spin connection. All degrees of freedom are encoded in the tetrad which, in the end, solves the symmetric and the antisymmetric part of the $f(T)$ field equations, and yields a vanishing torsion scalar in the Minkowski spacetime limit. We would like to stress that this is as equivalent as considering a non-vanishing spin connection and another tetrad, which together solve the symmetric and antisymmetric parts of the field equations \cite{Golovnev:2017dox, Hohmann:2017duq, Krssak:2015oua}.

\subsection{Covariant teleparallel gravity}
The fundamental variables in teleparallel theories of gravity are the tetrad of a Lorentzian metric $g = \eta_{ab} \theta^a \otimes \theta^b$, which can be expressed in local coordinates as 
\begin{align}
	\theta^a = h^a{}_{\mu}\dd x^\mu, \quad e_a = h_a{}^\mu \partial_\mu,\quad \theta^a(e_b) = \delta^a_b \quad \Rightarrow \quad g_{\mu\nu} = \eta_{ab}h^a{}_\mu h^b{}_\nu \,,
\end{align}
and a flat, metric compatible spin connection that is generated by local Lorentz matrices $\Lambda^a{}_b$
\begin{align}
	\omega^a{}_{b\mu} = \omega^a{}_{b\mu}(\Lambda) = \Lambda^a{}_c \partial_\mu (\Lambda^{-1})^c{}_{b},\quad \eta_{ab}\Lambda^a{}_c\Lambda^b{}_d = \eta_{cd}\,,
\end{align}
which poses torsion
\begin{align}
	T^a{}_{\mu\nu} = 2 \left(\partial_{[\mu}h^a{}_{\nu]} + \omega^a{}_{b[\mu} h^b{}_{\nu]}\right)\,.
\end{align}
In the Weitzenb\"ock gauge, the spin connection is set to be zero ($\omega^a{}_{b\mu}=0$) and so, the torsion tensor reduces to $T^a{}_{\mu\nu} = \partial_{[\mu}h^a{}_{\nu]}$. From here on we will work in the Weitzenb\"ock gauge. This is equivalent to having a non-vanishing spin-connection and a tetrad which solve their respective anti-symmetric part of the field equations. A detailed discussion about this equivalence can be found in Refs.~\cite{Golovnev:2017dox,Hohmann:2017duq,Krssak:2015oua}.

The teleparallel equivalent of General Relativity (TEGR) is constructed from the action
\begin{align}
	S_{\rm TEGR}= \int \dd^4x\ |h| \left( \frac{1}{2\kappa^2}  T + \mathcal{L}_{\rm m}(g,\Psi) \right)\,,\label{TEGR}
\end{align}
where $\kappa^2=8\pi$, $|h|=\det (h^a{}_\mu)=\sqrt{-g}$ is the determinant of the tetrad, $\mathcal{L}_{\rm m}(g,\Psi)$ is the matter Lagrangian for matter minimally coupled to gravity via the metric generated by the tetrads, and the so-called torsion scalar $T$ reads as follows
\begin{align}
	T =  T^a{}_{\mu\nu}S_a{}^{\mu\nu} =\frac{1}{2} \left(h_a{}^\sigma g^{\rho \mu} h_b{}^\nu + 2 h_b{}^\rho g^{\sigma \mu} h_a{}^\nu + \frac{1}{2} \eta_{ab} g^{\mu\rho} g ^{\nu\sigma} \right) T^a{}_{\mu\nu} T^b{}_{\rho\sigma}\,.
\end{align}
The superpotential $S_a{}^{\mu\nu}$ is given by $S_a{}^{\mu\nu} = \frac{1}{2}(K^{\mu\nu}{}_a - h_a{}^\mu T_\lambda{}^{\lambda\nu} + h_a{}^\nu T_\lambda{}^{\lambda\mu})$ in terms of the contortion tensor $K^{\mu\nu}{}_a = \frac{1}{2}(T^{\nu\mu}{}_a + T_a{}^{\mu\nu} - T^{\mu\nu}{}_a)$, and the appearing components of the metric are understood as a function of the tetrads.

The modified teleparallel theory of gravity we are investigating is $f(T)$ gravity, which is a straightforward generalisation of the action~\eqref{TEGR} as follows:
\begin{align}
	S_{f(T)} = \int \dd^4x\ |h| \left( \frac{1}{2\kappa^2}  f(T) + \mathcal{L}_{\rm m}(g,\Psi) \right)\,.
\end{align}
The function $f$ is an arbitrary function of the torsion scalar. Variation with respect to the tetrad $h^a{}_\mu$ yields the field equations~\cite{Krssak:2015oua}
\begin{align}\label{eq:fT}
	\frac{1}{4}f(T) h_a{}^\mu +  f_T\ \left( T^b{}_{\nu a} S_b{}^{\mu \nu } + \frac{1}{h}\partial_{\nu}(h S_a{}^{\mu \nu }) \right) +  f_{TT}\  S_a{}^{\mu\nu} \partial_\nu T &= \frac{1}{2}\kappa^2 \Theta_a{}^\mu\,,
\end{align}
with $\Theta_a{}^\mu$ being the energy-momentum tensor of the matter field, $f_T=\partial f/\partial T$ and  $f_{TT}=\partial^2 f/\partial T^2$. They can be  rewritten purely in terms of spacetime indices by contraction with $g_{\mu\rho}$ and $h^a{}_\sigma$ to take the form
\begin{align}
 H_{\sigma\rho} = \frac{1}{2}\kappa^2 \Theta_{\sigma\rho}\,.
\end{align}
Their symmetric part is sourced by the energy-momentum tensor, while their anti-symmetric part is a vacuum constraint for the matter models we consider. The latter is equal to the variation of the action with respect to the flat spin-connection components \cite{Golovnev:2017dox, Hohmann:2017duq},
\begin{align}
	 H_{(\sigma\rho)} = \frac{1}{2}\kappa^2 \Theta_{(\sigma\rho)}, \quad  H_{[\sigma\rho]} =0\,.
\end{align}
The explicit form of these equations can be found for example in Eqs.~(26) and (30) in \cite{Hohmann:2018rwf} by setting the scalar field $\phi$ to zero. We do not display these here since we will derive the spherically symmetric field equations directly from \eqref{eq:fT}.

\subsection{Spherical Symmetry in $f(T)$ gravity}\label{ref:ssecSphfT}
In this section, the $f(T)$ field equations for a spherically symmetric spacetime will be derived. Let us first start with the following spherically symmetric metric in standard spherical coordinates $(t,r,\theta, \phi)$:
\begin{equation}
ds^2=A \,dt^2-B\, dr^2-r^2(d\theta^2+\sin^2\theta d\phi^2)\,,\label{metric}
\end{equation}
where $A=A(r)$ and $B=B(r)$ are positive functions which depend on the radial coordinate. This means we consider the outside region of possible black holes. To calculate the field equations, we employ the following off-diagonal tetrad~\cite{Bohmer:2011si}
\begin{equation}
h^a{}_{\nu}=\left(
\begin{array}{cccc}
\sqrt{A} & 0 & 0 & 0 \\
0 & \sqrt{B} \cos (\phi ) \sin (\theta ) & r \cos (\phi ) \cos (\theta )  & -r \sin (\phi ) \sin (\theta )  \\
0 & \sqrt{B} \sin (\phi ) \sin (\theta )  & r \sin (\phi ) \cos (\theta )  & r \cos (\phi ) \sin (\theta ) \\
0 & \sqrt{B} \cos (\theta ) & -r \sin (\theta ) & 0 \\
\end{array}
\right)\label{tetrad}\,.
\end{equation}
This tetrad together with a vanishing spin connection consistently defines a spherically symmetric teleparallel geometry. Hence, it is consistent to derive the $f(T)$-field equations \eqref{eq:fT} from this tetrad with vanishing spin connection. Equivalently one could choose a diagonal tetrad with a non-vanishing spin connection~\cite{Hohmann:2019nat}, as was done in \cite{DeBenedictis:2016aze}.

For this setup the torsion scalar becomes
\begin{align}\label{scaletorsion}
  T=  -\frac{2 \left(\sqrt{B(r)}-1\right) \left(r A'(r)-A(r) \sqrt{B(r)}+A(r)\right)}{r^2 A(r) B(r)}\,.
\end{align}
Clearly, if $A\rightarrow 1$ and  $B\rightarrow 1$ (Minkowski limit), the torsion scalar vanishes.
Calculating the field equations~\eqref{eq:fT} contracted with $h^a{}_\sigma$ and an anisotropic fluid energy momentum-tensor, defined by the energy density $\rho = \Theta^0{}_0$, the radial and the lateral pressures $- p_{r} = \Theta^1{}_1 $ and $- p_{l} = \Theta^2{}_2 $ respectively, we find the non-vanishing independent spherically symmetric $f(T)$ field equations are the diagonal components $H^\mu{}_\mu$ (no sum taken),
\begin{eqnarray}
\frac{1}{2} \kappa ^2 \rho&=&\frac{r B(\sqrt{B}-1) A'+A (r B'+2 B^{3/2}-2 B)}{2 r^2 A B^2}f_T+\frac{(\sqrt{B}-1) }{r B}T' f_{TT}+\frac{1}{4} f \,,\label{Eq1}\\
\frac{1}{2} \kappa ^2 p_{r}&=& - \frac{r (\sqrt{B}-2) A'+2 A (\sqrt{B}-1) }{2 r^2 A B}f_{T} - \frac{1}{4} f\,,\label{Eq2}\\
\frac{1}{2} \kappa ^2 p_{l}&=&\frac{-r^2 B A'^2+r A \left(-r A' B'-4 B^{3/2} A'+2 B\left(r A''+3 A'\right)\right)+A^2 \left(-2 r B'-8 B^{3/2}+4 B^2+4 B\right)}{8 r^2 A^2 B^2}f_{T}\nonumber\\
&&+\frac{r A'-2 A (\sqrt{B}-1) }{4 r A(r) B(r)}T'f_{TT}-\frac{1}{4} f\,,\label{Eq3}
\end{eqnarray}
where primes denotes derivatives with respect to the radial coordinate. There are only three independent equations since $H^\phi{}_\phi \sim H^\theta{}_\theta$. This also demonstrates that the tetrad~\eqref{tetrad} is a so-called good tetrad, i.e., it solves the anti-symmetric field equations $H_{[\mu\nu]}=0$ with vanishing spin connection. We like to remark here that our choice of tetrad is not the only good tetrad in this sense. The tetrad presented in \cite{Tamanini:2012hg} could be chosen equally well from this point of view. However not all of these good tetrads, in the sense of the field equations, yield consistently a torsion scalar that vanishes for the Minkowski spacetime limit.

A consistency check of the field equations is to set $f(T) = T$, hence $f_{T} = 1$ and $f_{TT}=0$, and to see if for $\rho=p_r=p_l=0$, Schwarzschild geometry ($A(r) = B(r)^{-1} = 1 - 2M/r$) solves the field equations, which is the case.

The last remark leads us to the point that the equations we derived above differ from the spherically symmetric $f(T)$ field equations in \cite{Ruggiero:2015oka}. First, their field equations~(8)-(10) are not solved by Schwarzschild geometry for $f(T)=T$ and $\rho=p_r=p_l=0$, as it is the case for our equations. Second the off-diagonal tetrad they choose has the problem that its torsion scalar does not vanish for the Minkowski spacetime limit, which means that it is not the correct tetrad to which one associates a vanishing spin connection.

We will now solve the $f(T)$ field equations for a specific power law choice for $f$ to first order in a perturbation around Minkowski and Schwarzschild spacetime geometry. In each instance we will compare our results and the ones obtained in \cite{Ruggiero:2015oka} and \cite{DeBenedictis:2016aze}.

\section{Weak power law $f(T)$ gravity} \label{sec:weak}
In this section we turn our focus to a power-law $f(T)$ model that reads as
\begin{equation}\label{eq:weakfT}
f(T)=T+\frac{1}{2}\epsilon\,\alpha T^p\,,
\end{equation}
where $\alpha$ and $p$ are constants and $\epsilon\ll 1$ is a small tracking parameter that, similarly as it was done in~\cite{DeBenedictis:2016aze}, is used to make the series expansion in a coherent way. It allows us to easily track quantities that are considered to be small throughout our calculations.

Since we are interested in perturbations  of a Schwarzschild background, the ansatz we employ for the metric coefficients is
\begin{align}
A(r)&=1-\frac{2M}{r}+\epsilon \, a(r)\,,\\  
B(r)&= \left(1-\frac{2M}{r} \right)^{-1}+\epsilon \,  b(r)\,,\label{expansion}
\end{align}
where $a(r)$ and $b(r)$ are functions of the radial coordinate. If one uses the above metric coefficients in the $f(T)$ power-law spherically symmetric field equations~\eqref{Eq1}--\eqref{Eq3} and then expands up to first order in $\epsilon$, the equations become 
\begin{eqnarray}
\frac{1}{2} \kappa ^2\epsilon \rho&=& \epsilon \left( \alpha \frac{(-1)^{p+1}\ 2^{p-3}\ (p-1)}{(r^2 \mu)^p}\ (\mu-1)^{2p-1}\left(\mu-1+p(1+\mu(2+5\mu)) \right) - \frac{\mu^2 (\mu^2 - 2) b(r)}{2 r^2}  + \frac{\mu^4 b'(r)}{2 r}\right)\label{Eq1b}\,,\\
\frac{1}{2}\kappa^2 \epsilon p_{r}&=&\epsilon \left( - \alpha \frac{(-1)^{p+1}\ 2^{p-3}\ (p-1)}{(r^2 \mu)^p} + \frac{ a'(r)}{2 r} + \frac{(\mu ^2-1) a(r)}{2 \mu ^2 r^2} - \frac{\mu ^2 b(r)}{2 r^2} \right)\,,\label{Eq2b}\\
\frac{1}{2}\kappa^2 \epsilon p_{l}
&=& \epsilon \left(  - \alpha \frac{(-1)^{p+1}\ 2^{p-5}\ (p-1)}{\mu (r \mu)^{p}}\ (\mu-1)^{2p} (p+2(2+p)\mu+5p \mu^2)\right. \nonumber \\
&+&\left.\frac{1}{4} a''(r) + \frac{\left(3 \mu ^2-1\right)a'(r)}{8 \mu ^2 r} - \frac{\left(\mu ^4-1\right)  a(r)}{8 \mu ^4 r^2} - \frac{\mu^2 \left(\mu ^2+1\right) b'(r)}{8 r} + \frac{\left(\mu ^4-1\right) b(r)}{8 r^2} \right)\,,\label{Eq3b}
\end{eqnarray}
where $\mu=(1-2M/r)^{1/2}$ was introduced for simplicity and we assumed that the energy momentum-tensor is zero to zeroth order in $\epsilon$. The latter assumption is necessary in order to have Schwarzschild geometry as consistent zeroth order solution. As usual in perturbation theory at this stage, the small parameter $\epsilon$ drops out of the equations and they can be solved for the first order perturbations $a(r)$ and $b(r)$. Eq.~\eqref{Eq2b} is an algebraic equation for $b(r)$ that can be easily solved, yielding 
\begin{eqnarray}
b(r)&=& \alpha \frac{(-1)^p\ 2^{p-2}\ (p-1) r^2 (\mu-1)^{2p}}{\mu^2 (r^2 \mu)^p} + \frac{(\mu ^2-1) a(r)}{\mu ^4}  + \frac{r a'(r)}{\mu^2} - \frac{\kappa ^2 r^2 }{\mu ^2 }p_r \label{bis}\,.
\end{eqnarray}
Inserting this result for $b(r)$ in \eqref{Eq1b} and~\eqref{Eq3b}, while setting $p_r = p_l$, we obtain one remaining partial differential equation, which can be solved for $a(r)$
\begin{equation}
    a''+\frac{2 a'}{r}-\frac{\alpha  2^{p-3} \left(4 (\mu -1) \mu ^2+\left(5 \mu ^3+7 \mu ^2+3 \mu +1\right) p^2-\left(9 \mu ^3+3 \mu ^2+3 \mu +1\right) p\right) r^{-3 p} \left(-\frac{(\mu -1)^2 r}{\mu }\right)^p}{(\mu -1) \mu ^2}=0\,,
\end{equation}
In order to continue solving the equations, we will separate the study into two branches: A) $M=0$, and $p_r=p_{l}=-\rho=-\Lambda$ as it was studied in Ref.~\cite{Ruggiero:2015oka}, with the additional constraint $p > 1/2$ to guarantee well-defined field equations. We redo the calculations of \cite{Ruggiero:2015oka}, since we find a completely different result for the $M=0$. B) $M\neq0$, $\rho=p_r=p_l=0$ and $p=2$ to $10$. For $p=2$ we reproduce the result from \cite{DeBenedictis:2016aze}.

\subsection{Minkowski background ($M=0$)}
If one assumes $p>1/2$, perturbations around a Minkowski background can be studied by setting $M=0$ ($\mu=1$). It can immediately be seen from Eqs.~\eqref{Eq1b}-\eqref{Eq3b} that the influence of the teleparallel perturbation (e.g. the terms proportional to $\alpha$) drops out. As a consequence one obtains the usual first order (A)dS Schwarzschild spacetime geometries as solutions of the perturbed field equations with a cosmological constant as first order matter source, i.e., $p_r=p_{l}=-\rho=-\Lambda$. The perturbation functions $a(r)$ and $b(r)$ are easily determined from \eqref{Eq1b}--\eqref{Eq3b}. The metric coefficients $A(r)$ and $B(r)$ become
\begin{align}
A(r)&=1 + \epsilon \left(C_2 - \frac{C_1}{r} - \frac{1}{3}\kappa ^2 \Lambda  r^2 \right)\,,\label{A}\\
B(r)&=1 + \epsilon \left(\frac{C_1}{r} + \frac{1}{3}\kappa ^2 \Lambda  r^2 \right)\,,\label{B}
\end{align}
for all $p > 1/2$. Here $C_1$ and $C_2$ are integration constants labelling the linearised Schwarzschild solution of general relativity and a constant shift of the Minkowski metric respectively. Usually they are chosen to be $C_1 = 2M$ and $C_2 = 0$. The $\Lambda$ term appears due to a non-vanishing cosmological constant we assumed as a first order matter source.

The solutions we find are completely different to the ones presented in \cite{Ruggiero:2015oka}. The source of this discrepancy lies in our choice of the tetrad \eqref{tetrad}, to which we associate a vanishing spin connection. The tetrad chosen in the previous work had the drawback that its torsion scalar, see Eq.~(9) in \cite{Ruggiero:2015oka}, does not vanish in the Minkowski spacetime limit $A\to 1$, $B\to 1$, but gives $8/r^2$. In turn this leads to an infinite action for Minkowski spacetime. As we mentioned earlier, our tetrad avoids this complication by having a vanishing torsion scalar for the Minkowski spacetime limit.

\subsection{Schwarzschild background $M\neq0$}\label{ssec:schw}
In this section, we will focus our study on perturbations of Schwarzschild geometry ($M\neq0$) induced by weak power-law $f(T)$ gravity
\begin{eqnarray}
f(T)=T+\frac{1}{2}\alpha \epsilon\, T^p\,,
\end{eqnarray}
exemplary for $p=2$ to $p=4$, since a solution for general $p$ cannot be obtained. For higher integer values of $p$ the solutions have a similar form, but it is not further insightful to display the long expressions.

The general structure of the vacuum solutions,i.e.\ $\rho=p_r=p_l=0$,  for all $p$ is
\begin{align}
    A(r)&= 1 - \frac{2M}{r} +\epsilon\left(-\frac{C_1}{r}+C_2 + \alpha \bar a(r)\right)\,,\label{astruct}\\
    B(r)&= \frac{1}{1 - \frac{2M}{r}} +\epsilon\left(\frac{\left(\frac{C_1}{r}-\frac{2 C_2 M}{r}\right)}{\left(1-\frac{2 M}{r}\right)^{2}} + \alpha \bar b(r)\right)\,,\label{bstruct}\,.
\end{align}
The integration constants $C_1$ and $C_2$ are determined in a power series expansion in $\frac{1}{r}$, such that the zeroth and first order terms in this expansion vanish. Physically this means we use the integration constants to avoid an influence of the teleparallel perturbation in the weak field, respectively, large distance, limit. After integration constants have been found, the solutions take the form
\begin{align}
    A(r)&= 1 - \frac{2M}{r} +\epsilon\alpha \hat a(r)\,,\label{astruct2}\\
    B(r)&= \frac{1}{1 - \frac{2M}{r}} +\epsilon \hat b(r)\,.\label{bstruct2}
\end{align}

For $p=2$ we reproduce the solutions found in \cite{DeBenedictis:2016aze}
\begin{eqnarray}
    A(r)&=& 1- \frac{2M}{r}+\epsilon\left(-\frac{C_1}{r}+C_2-\alpha\left[\frac{M^2+6 M r+r^2}{M r^3}-\frac{16 \left(1-\frac{2 M}{r}\right)^{3/2}}{3 M^2}+\frac{(1-\frac{3 M}{r})}{2 M^2} \ln \left(1-\frac{2M}{r}\right)\right]\right)\,,\label{afin}\\
    B(r)&=& \frac{1}{1- \frac{2M}{r}} + \epsilon\left(\frac{\left(\frac{C_1}{r}-\frac{2 C_2 M}{r}\right)}{\left(1-\frac{2 M}{r}\right)^{2}} -\alpha\left[-\frac{8 (3 M^2-7 M r+2 r^2)}{3 M r^3 (1-\frac{2 M}{r})^{3/2}} +\frac{25  M-23 r}{r^3 (1-\frac{2 M}{r})^2}+\frac{\ln\left(1-\frac{2M}{r}\right)}{2Mr(1-\frac{2M}{r})^{2}}\right]\right)\,,\label{bfin}
\end{eqnarray}
which can be be expressed conveniently in terms of the variable $\mu=(1-2M/r)^{1/2}$, giving
\begin{eqnarray}
    A(r)&=& \mu^2 +\epsilon\left(- \frac{C_1}{r}+C_2 - \alpha \frac{51-93 \mu^2 -128 \mu^3 + 45 \mu^4 - 3 \mu^6 - 12 (1-3 \mu^2) \ln(\mu)}{6 r^2 (1 - \mu^2)^2}\right)\,,\label{afinX}\\
    B(r)&=& \frac{1}{\mu^2}+ \epsilon\left(\frac{\left(\frac{C_1}{r} - (1 - \mu^2)C_2\right)}{\mu^4} + \alpha \frac{63 -24 \mu + 12 \mu^2 +64 \mu^3 - 75 \mu^4 + 24 \mu^5 - 12 \ln(\mu)}{6 r^2 \mu^4 (1 - \mu^2)}\right)\,.\label{bfinX}
\end{eqnarray}
Determining the integration constants from the $\frac{1}{r}$ expansions 
\begin{align}
    A(r) \sim \left(\frac{16 \alpha}{3 M^2} + C_2 \right) - \left(\frac{16 \alpha}{M}+ C_1\right)\frac{1}{r} + \mathcal{O}\Big(\frac{1}{r^2}\Big),\quad 
    B(r) \sim \left(\frac{16 \alpha}{3M} + C_1 - 2 M C_2\right)\frac{1}{r} + \mathcal{O}\Big(\frac{1}{r^2}\Big)
\end{align} 
yields $C_2 = -16\alpha/(3 M^2)$ and $C_1 = -16 \alpha/M$. For easy comparison with previous approaches we display the solutions \eqref{afinX} and \eqref{bfinX} one more time with this choice of integration constants
\begin{align}
    A(r)&= \mu^2 + \epsilon\alpha \frac{ 13 - 99 \mu^2 + 128 \mu^3 - 45 \mu^4 + 3 \mu^6 + 12(1-3\mu^2) \ln(\mu)}{6 r^2 (1 - \mu^2)^2}\,,\label{afinX2}\\
    B(r)&= \frac{1}{\mu^2} - \epsilon \alpha \frac{1 + 24 \mu - 12 \mu^2 - 64 \mu^3 + 75 \mu^4 - 24 \mu^5 + 12 \ln(\mu)}{6 r^2 \mu^4 (1-\mu^2)}\,.\label{bfinX2}
\end{align}
The leading order terms for the torsion scalar \eqref{scaletorsion} for the weak squared power-law case in a Schwarzschild background behaves as
\begin{align}\label{eq:Tscalarsol}
    T = -\frac{2 (\mu -1)^2}{\mu  r^2} + \epsilon \alpha \frac{13 - 36 \mu + 108 \mu^2 - 184 \mu^3 + 135 \mu^4 - 36 \mu^5 + 12 \ln{(\mu)}}{6 r^4 \mu^3}\,.
\end{align}
For the specific choice of $C_1$ and $C_2$ we find that in the $M\to 0$ limit $A(r) \to 1$, $B(r)\to 1$, and $T\to 0$. This result coincides with the one presented in \cite{DeBenedictis:2016aze}.

For $p=3$ and $p=4$ we find similar solutions that can be expressed as
\begin{align}
    A(r) = \mu^2 + \epsilon \,\alpha\, \hat a(r)\,,\quad B(r) = \frac{1}{\mu^2} + \epsilon \,\alpha\, \hat b(r)\,,\quad T(r) = -\frac{2 (\mu -1)^2}{\mu  r^2} + \epsilon\, \alpha\, \hat T(r)\,.
\end{align}

\begin{itemize}
    \item $p=3$
      \begin{subequations}\label{eq:p3}
    \begin{align}
        \hat a(r) &= \bigg[ -280 \mu^{12} + 945 \mu^{11} + 1120 \mu^{10} - 8295 \mu^9 + 6984 \mu^8 + 18060 \mu^7 - 37632 \mu^6 + 1260 \mu^5 + 86520 \mu^4\nonumber\\
        & - 62909 \mu^3 - 10080 \mu^2 - 7560 \left(7 \mu^2-1\right) \mu \log (\mu) + 178\mu + 2520 \bigg] \left(315 r^4 \mu \left(\mu^2-1\right)^4\right)^{-1}\,,\\
        \hat b(r) &= \bigg[ 3780 \mu^{11}-19040 \mu^{10}+27405 \mu^9+16560 \mu^8-81480 \mu^7+56448 \mu^6+44100 \mu^5-77280 \mu^4+23940 \mu^3\nonumber\\
        &+10080 \mu^2-9553 \mu+7560 \mu \log (\mu)+5040\bigg] \left(315 r^4 \mu^5 \left(\mu^2-1\right)^3\right)^{-1} \,,\\
        \hat T(r) &= \bigg[7(\mu-1) \big(6300 \mu^{10}-30380 \mu^9+44905 \mu^8+8545 \mu^7-88055 \mu^6+74233 \mu^5+12493 \mu^4-49667 \mu^3\nonumber\\
        &+27193 \mu^2-10607 \mu-2520\big)+7560 \mu \log (\mu)\bigg]\left(315 r^6 \mu^4\left(\mu^2-1\right)^2\right)^{-1}\,.
    \end{align}
      \end{subequations}
    \item $p=4$
    
\begin{subequations}\label{eq:p4}
    \begin{align}
        \hat{a}(r) &=  -4 \bigg[-6435 \mu^{18}+36960 \mu^{17}-32175 \mu^{16}-221760 \mu^{15} + 552552 \mu^{14} +145600 \mu^{13}-1963962 \mu^{12}\nonumber\\
        & +1693120 \mu^{11} + 2642640 \mu^{10}-5436288 \mu^9+330330 \mu^8 + 7495488 \mu^7-6846840 \mu^6-4804800 \mu^5\nonumber\\
        &+3912986 \mu^4 +2882880 \mu^3 +55139 \mu^2 + 720720\left(11 \mu^2-1\right) \mu^2 \log (\mu)-480480 \mu\nonumber\\
        &+45045\bigg] \left( 15015 r^6 \mu^2 \left(\mu^2-1\right)^6\right)^{-1}\,,\\
       \hat{b}(r) &=  - 4 \bigg[120120 \mu^{17}-842985 \mu^{16}+1940400 \mu^{15}-60060 \mu^{14}-7141680 \mu^{13}+10198188 \mu^{12}+3443440 \mu^{11}\nonumber\\
        &-20810790 \mu^{10}+13590720 \mu^9+12222210 \mu^8-20612592 \mu^7+4504500 \mu^6+8888880 \mu^5-6666660 \mu^4\nonumber\\
        &+720720 \mu^3+1451534 \mu^2-720720 \mu^2 \log (\mu)-1081080 \mu+135135\bigg] \left(15015 r^6 \mu^6 \left(\mu^2-1\right)^5\right)^{-1}\,,\\
        \hat{T}(r) &= 2 \bigg[1441440 \mu^2 \log (\mu)-2 (\mu-1) \Big(210210 \mu^{16}-1443585 \mu^{15}+3379695 \mu^{14}-1004685 \mu^{13}-9227445 \mu^{12} \nonumber\\
        &+15024783 \mu^{11}-270497 \mu^{10}-21081287 \mu^9+18275173 \mu^8+4731643 \mu^7-15880949 \mu^6+7362271 \mu^5\nonumber\\
        & + 2197111\mu^4-3388469 \mu^3+1656571 \mu^2+225225 \mu-45045\Big)\bigg]\left(15015 r^8 \mu^5 \left(\mu^2-1\right)^4\right)^{-1}\,.
    \end{align}
    \end{subequations}
    \end{itemize}
For the other higher values of $p$ the calculation follows the same scheme and their solutions are similar to the above ones. Since they are lengthy and not insightful, we will not present them explicitly. 


\section{Particle motion Phenomenology}\label{sec:pheno}
To relate the influence of a teleparallel modification of general relativity to observables, we study the motion of test particles in the background solution defined by the metric coefficients \eqref{afin} and \eqref{bfin}. We explicitly derive the photon sphere around the black hole and the perihelion shift of nearly circular orbits. Nowadays the photon sphere is of particular interest since it defines the edge of the shadow of a black hole while the perihelion shift was already derived in \cite{DeBenedictis:2016aze}, but on the basis of an erroneous solution, as we discussed above.

\subsection{Geodesic equation and effective potential}
The worldline $q(\tau)$ of test particles in a curved spacetime is determined by the Euler-Lagrange (EL) equations
\begin{equation}\label{4} 
\frac{d}{d\tau}\left(\frac{\partial \mathcal{L}}{\partial \dot{q}^\mu}\right)-\frac{\partial \mathcal{L}}{\partial q^\mu}=0\,, 
\end{equation}
of the Lagrangian
\begin{equation}\label{7}
2\mathcal{L}=g_{\mu\nu}\dot{q}^{\mu}\dot{q}^{\nu}=\left(1-\frac{2 M}{r}+ \epsilon\, a(r)\right)\dot{t}^2-\left(\frac{1}{1-\frac{2 M}{r}}+ \epsilon\, b(r)\right)\dot{r}^2-r^2\dot{\theta}^2-r^2\sin^2{\theta}\dot{\phi}^2\,,
\end{equation}
where $q^{\mu}(\tau)=(t(\tau), x(\tau), \theta(\tau), \phi(\tau))$,  and $\dot{q}^\mu$ denotes the derivative of $q^\mu$ with respect to the affine parameter $\tau$. The perturbation functions $a(r)$ and $b(r)$ under consideration can be read off in \eqref{afin} and \eqref{bfin}.

To solve the EL equations we employ the usual scheme for spherically symmetric spacetimes: we restrict ourselves to motion in the equatorial plane and set $\theta=\pi/2$, and we derive the usual conserved quantities the energy $k$ and angular momentum $h$
\begin{align}
	k &= \frac{\partial \mathcal{L}}{\partial \dot t} = \left(1-\frac{2 M}{r}+ \epsilon\, a(r)\right)\dot{t}\,,\\
	h &= \frac{\partial \mathcal{L}}{\partial \dot \phi} = r^2 \dot \phi\,.
\end{align}
We obtain the effective potential to first order in $\epsilon$ from the constancy of the Lagrangian, expressed in terms of the conserved quantities
\begin{align}\label{17}
	\left(1 - \epsilon \frac{a(r)}{1-\frac{2 M}{r}}\right)\frac{k^2}{1-\frac{2 M}{r}}-\left(\frac{1}{1-\frac{2 M}{r}}+ \epsilon b(r)\right)\dot{r}^2 - \frac{h^2}{r^2} + \mathcal{O}(\epsilon^2) = \sigma\,,
\end{align}
where $\sigma = 0$ for massless particles and $\sigma = 1$ for massive particles. For further calculations we rearrange Eq.~\eqref{17} as
\begin{align}
	0 &= \frac{1}{2}\dot r^2 - \frac{1}{2} k^2 + \frac{1}{2} \frac{h^2}{r^2}\left(1-\frac{2M}{r}\right) + \frac{1}{2} \sigma \left(1-\frac{2M}{r}\right) \nonumber\\
	&+ \frac{\epsilon}{2} \left[ k^2 \left( \frac{a(r)}{1-\frac{2M}{r}} + b(r)\left(1-\frac{2M}{r}\right)\right) - b(r) \frac{h^2}{r^2}\left(1-\frac{2M}{r}\right)^2 - \sigma b(r) \left(1-\frac{2M}{r}\right)^2\right]\,,
\end{align}
so we can read off the effective potential to first order in $\epsilon$
\begin{align}\label{eq:pot}
	V(r) &= - \frac{1}{2} k^2 + \frac{1}{2} \left(1-\frac{2M}{r}\right) \left( \frac{h^2}{r^2} + \sigma \right) \nonumber\\
	&+ \frac{\epsilon}{2} \left[ k^2 \left( \frac{a(r)}{1-\frac{2M}{r}} + b(r)\left(1-\frac{2M}{r}\right)\right) - b(r)\left( \sigma + \frac{h^2}{r^2} \right)\left(1-\frac{2M}{r}\right)^2\right]
\end{align}
from
\begin{align}
	\frac{1}{2}\dot r^2 + V(r)=0\,.
\end{align}
For the analysis of the perihelion shift, it is necessary to reparametrise $r(\tau)$ as $r(\phi)$, which amounts to the equation
\begin{align}\label{eq:rphi}
	\frac{1}{2}\frac{\dot r^2}{\dot \phi ^2} + \frac{1}{\dot \phi^2}V(r)  = \frac{1}{2}\left(\frac{dr}{d\phi}\right)^2 + \frac{r^4}{h^2}V(r)=0\,.
\end{align}

\subsection{Photon sphere and perihelion shift}\label{ssec:phsph}
For circular orbits (e.g $r=\textrm{const.},\ \dot r=0$) the effective potential and its derivative have to vanish. We perturb the radial coordinate of the circular orbit $r_c = r_0+\epsilon\, r_1$, the angular momentum $h = h_0 + \epsilon \,h_1$, and the energy $k = k_0 + \,\epsilon k_1$ and solve both equations $V = 0$ and $V' = 0$ order by order. 
For circular photon orbits, $\sigma =  0$, solving the zeroth order equations yields
\begin{align}
	r_0 = 3M,\quad h_{0\pm} = \pm 3 \sqrt{3} k_0 M
\end{align} 
and the first order perturbation gives, for the different values of $p$, the following numerical values:
\begin{align}
    (p=2) \quad r_1 &\approx 14133.8000 \cdot 10^{-6} \frac{\alpha}{M}\,,    && 
    (p=3) \quad r_1 \approx -1362.5400 \cdot 10^{-6} \frac{\alpha}{M^3}\,,  \\
    (p=4) \quad r_1 &\approx 121.3220 \cdot 10^{-6} \frac{\alpha}{M^5}\,,    && 
    (p=5) \quad r_1 \approx -10.2582 \cdot 10^{-6} \frac{\alpha}{M^7}\,,    \\
    (p=6) \quad r_1 &\approx 8.3757 \cdot 10^{-6} \frac{\alpha}{M^9}\,,      &&
    (p=7) \quad r_1 \approx -0.6670 \cdot 10^{-6} \frac{\alpha}{M^{11}}\,,  \\
    (p=8) \quad r_1 &\approx 0.0521 \cdot 10^{-6} \frac{\alpha}{M^{13}}\,,   &&
    (p=9) \quad r_1 \approx -0.0040 \cdot 10^{-6} \frac{\alpha}{M^{15}}\,,  \\
    (p=10)\quad r_1 &\approx 0.0003 \cdot 10^{-6} \frac{\alpha}{M^{17}}\,.
\end{align}

We clearly see that for positive $\alpha$ and even $p$ the teleparallel perturbation of general relativity yields a larger photon sphere around a spherically symmetric black hole and thus predicts a larger black hole shadow. For odd $p$ a smaller shadow is predicted. Moreover it is evident that the larger $p$ the smaller the first order influence of the teleparallel perturbation. The relation between the photon sphere and teleparallel perturbations of general relativity is investigated here for the first time.

For circular timelike orbits, $\sigma =  1$ it is also possible to solve the equations $V = 0$ and $V' = 0$. However the appearing expressions are not very insightful. The important observation is that teleparallely perturbed general relativity, not surprisingly, allows for circular orbits, which will be the starting point for the derivation of the perihelion shift now. We consider a perturbation around a circular orbit $r_c$ and plug the ansatz $r(\phi)= r_{c} + r_\phi(\phi)$ into \eqref{eq:rphi} and obtain
\begin{align}
    \left(\frac{d r_\phi}{d\phi}\right)^2 = - 2 \frac{(r_c + r_\phi)^4}{h^2} V(r_c + r_\phi)\,.
\end{align}
Assuming that the ratio $r_\phi/r_c$ is small, the right-hand side can be expanded into powers of this parameter to second order
\begin{align}
    \left(\frac{d r_\phi}{d\phi}\right)^2 
    = - \frac{r_c^4}{h^2} V''(r_c)r_\phi^2 + \mathcal{O}\left(\tfrac{r_\phi^3}{r_0^3}\right) \,,
\end{align}
where we used that for circular orbits $V(r_c) = 0$ and $V'(r_c)=0$, as discussed above. The solution $r_\phi$ thus oscillates with the wave number $K = \sqrt{\frac{r_c^4}{h^2} V''(r_c)}$ and the perihelion shift is given by
\begin{align}
    \Delta \phi =2\pi\Big(\frac{1}{K}-1\Big) =2\pi \left(\frac{h}{r_c^2\sqrt{V''(r_c)}}-1\right)\,.
\end{align}
To derive the explicit expression for the perihelion shift for massive objects we consider the potential $V$ with $\sigma=1$, see \eqref{eq:pot}, its first derivative $V'$ and its second derivative $V''$. We evaluate the equations $V(r_c) = 0$ and $V'(r_c) = 0$ with $h= h_0 + \epsilon\, h_1$ and $k = k_0 +\epsilon\, k_1$. The zeroth order of these equations determines $h_0(r_c)$ and $k_0(r_c)$ as
\begin{align}
    h_{0\pm} = \pm \frac{\sqrt{M}r_c}{r_c-3M},\quad k_{0\pm} = \pm \frac{2M-r_c}{\sqrt{r_c(r_c-3 M)}}\,.
\end{align}
The first order determines $h_1(r_c)$ and $k_1(r_c)$. Depending on the choice of the sign of $h_0$ we obtain two different solutions for $h_1$ (the sign of $k_0$ is irrelevant here)
\begin{align}
    h_{1\pm} = \mp \frac{r_c^2(2 M a(r_c) - r_c (r_c - 2M) a'(r_c))}{4 \sqrt{M}\sqrt{r_c - 3M}^3}\,.
\end{align}
The sign labeling $h_{1\pm}$ refers to the sign chosen of the zeroth order $h_{0\pm}$, which was chosen to calculate $h_{1\pm}$. There is no need to derive $k_1$ explicitly, since it does not enter the perihelion shift equation. Having obtained the constants of motion for the circular orbit we can derive the perihelion shift by plugging the values into $V''(r, k_0, h_0, h_1)$ to obtain $V''(r_c)$ alone. Due to the different solutions for the constants of motion there exist two options to derive the perihelion shift
\begin{align}
    \Delta\phi(h_{0+},h_{1+})\,, \quad \Delta\phi(h_{0-},h_{1-})\,,
\end{align}
which are related to each other through
\begin{align}
    \Delta\phi(h_{0-},h_{1-}) &=  - 4\pi -  \Delta\phi(h_{0+},h_{1+})\,.
\end{align}
Expanding the perihelion shift into a power series in the variable $q = \frac{M}{r_c}$ yields
\begin{align}\label{eqn:perihelionshift}
    \Delta\phi(h_{0+},h_{1+}) &= 6 \pi q + 27 \pi q^2 + \mathcal{O}(q^3) + \epsilon\,\,\pi \hat\Delta\phi_{p}  + \mathcal{O}(\epsilon^2)\,,
\end{align}
$\hat\Delta\phi_{p}$ is the leading order teleparallel perturbation of the usual GR result. For the different values of $p$ it is given by
\begin{align}\label{eqn:contribution}
\hat\Delta\phi_{p=2}= & \frac{8 q^2}{r_c^2} \,, &&
\hat\Delta\phi_{p=3}= - \frac{48 q^4}{r_c^4}\,, &&
\hat\Delta\phi_{p=4}= \frac{192 q^6}{r_c^6}\,, && 
\hat\Delta\phi_{p=5}= -\frac{640 q^8}{r_c^8}\,, &&
\hat\Delta\phi_{p=6}= \frac{1920 q^{10}}{r_c^{10}}\,, \nonumber\\
\hat\Delta\phi_{p=7}= &-\frac{5376 q^{12}}{r_c^{12}}\,, &&
\hat\Delta\phi_{p=8}= \frac{14336 q^{14}}{r_c^{14}}\,, &&
\hat\Delta\phi_{p=9}= -\frac{36864 q^{16}}{r_c^{16}}\,, &&
\hat\Delta\phi_{p=10}= \frac{92160 q^{18}}{r_c^{18}}\,.
\end{align}
The qualitative behaviour of the perihelion shift is always the same, only the numerical values differ. As for the photon sphere, the higher $p$, the smaller the influence of the teleparallel perturbation and corrections to the perihelion shift appear only in higher orders in $q$.

Since the influence of the teleparallel perturbation decreases with higher power in $p$ the most strict bound on $\alpha$ is obtained for $p=2$ and is the one obtained in \cite{DeBenedictis:2016aze}. 

We expect to be able to find stronger bounds from the upcoming study on teleparallel perturbations of rotating black holes.

\section{Conclusion}\label{sec:conclusion}

In this paper we presented the first order influence of a teleparallel power law $f(T)$ gravity perturbation of general relativity, in spherical symmetry. The central results of this article are as follows:
\begin{itemize}
    \item To first order, a power law perturbation of the type $f(T) = T + \frac{\alpha}{2}T^p$ yields no teleparallel correction to Minkowski spacetime for $p>1/2$.
    \item The explicit derivation of the first order teleparallel $f(T) = T + \frac{\alpha}{2}T^p$ corrections to Schwarzschild geometry for $p=2$ to $p=10$, displayed in Eqs.~\eqref{afinX} and \eqref{bfinX} for $p=2$ and in \eqref{eq:p3} and \eqref{eq:p4} for $p=3$ and $p=4$, respectively. The perturbed solutions for higher power-law parameter $p$ have the same structure but they are lengthy and for this reason, we do not present them.
\end{itemize}
The latter allowed us to calculate the teleparallel modifications of the photon sphere and the perihelion shift: two observables which are experimentally accessible and can be used to check the viability of $f(T)$ models. For both observables we find that the larger $p$, the smaller the influence of the teleparallel modification. Thus, the $p=2$ model is most constraint from the perihelion shift of mercury, which is $|\alpha|_{\rm max}=2.20\cdot 10^{20} \,\mathrm{km}^{2}$ according to \cite{DeBenedictis:2016aze}. We expect to find further, stronger constraints, for the different models by studying rotating black holes and their phenomenology.

The results we presented improve and extend existing results on first order power law $f(T)$-models, which were presented in~\cite{Ruggiero:2015oka} and \cite{DeBenedictis:2016aze}. In the first article the tetrad chosen was not compatible with a vanishing spin connection and the field equations were incorrect. During our derivations, we paid particular attention to present all necessary steps in the perturbation theory, so that our results are easily reproducible. 

An important opportunity, that our presented approach here offers is to investigate the connection between the vanishing of the first order contributions of teleparallel corrections around Minkowski spacetime and the non-vanishing of the corrections around Schwarzschild geometry, with the degrees of freedom of $f(T)$ gravity; the latter being debated in the literature~\cite{Golovnev:2018wbh,Ferraro:2018axk,Ferraro:2018tpu}. Here we considered static perturbations; in the future, non-static spherically and axially symmetric gravitational waves from weak power law $f(T)$ gravity will be investigated and complement the gravitational wave analysis of $f(T)$ gravity around Minkowski and FLRW geometry~\cite{Farrugia:2018gyz,Golovnev:2018wbh,Nunes:2018evm} and also at the astrophysical level with compact binary coalescences~\cite{Nunes:2019bjq}. 

This paper is a first step towards a complete phenomenological catalogue of observables, which shall be derived in weak power law $f(T)$-gravity to systematically check its consistency with observations. The next step in this program is to consider axially symmetric perturbations around Kerr spacetime, to derive the change in the photon regions, which will have an imprint on the predictions of the shape of the black hole's shadow.

\begin{acknowledgments}
The authors would like to thank Jackson Levi Said, Gabriel Farrugia, Andrew DeBenedictis and Sasa Ilijic for very fruitful discussions. CP was supported by the Estonian Research Council and the European Regional Development Fund through the Center of Excellence TK133 ``The Dark Side of the Universe''. KF gratefully acknowledges support by the DFG, within the Research Training Group \textit{Models of Gravity} and mobility funding from the European Regional Development Fund through \textit{Dora Plus}. SB is supported by Mobilitas Pluss N$^\circ$ MOBJD423 by the Estonian government. 
\end{acknowledgments}

\appendix

\bibliographystyle{utphys}
\bibliography{PFTS}

\providecommand{\href}[2]{#2}\begingroup\raggedright\begin{thebibliography}{10}

\bibitem{Abbott:2016blz}
{\bf LIGO Scientific, Virgo} Collaboration, B.~P. Abbott {\em et al.},
  ``{Observation of Gravitational Waves from a Binary Black Hole Merger},''
  \href{http://dx.doi.org/10.1103/PhysRevLett.116.061102}{{\em Phys. Rev.
  Lett.} {\bf 116} (2016) no.~6, 061102},
\href{http://arxiv.org/abs/1602.03837}{{\tt arXiv:1602.03837 [gr-qc]}}.

\bibitem{Akiyama:2019cqa}
{\bf Event Horizon Telescope} Collaboration, K.~Akiyama {\em et al.}, ``{First
  M87 Event Horizon Telescope Results. I. The Shadow of the Supermassive Black
  Hole},'' \href{http://dx.doi.org/10.3847/2041-8213/ab0ec7}{{\em Astrophys.
  J.} {\bf 875} (2019) no.~1, L1},
\href{http://arxiv.org/abs/1906.11238}{{\tt arXiv:1906.11238 [astro-ph.GA]}}.

\bibitem{Aldrovandi:2013wha}
R.~Aldrovandi and J.~G. Pereira,
  \href{http://dx.doi.org/10.1007/978-94-007-5143-9}{{\em {Teleparallel
  Gravity}}}, vol.~173.
\newblock Springer, Dordrecht,
2013.
\newblock

\bibitem{Cai:2015emx}
Y.-F. Cai, S.~Capozziello, M.~De~Laurentis, and E.~N. Saridakis, ``{f(T)
  teleparallel gravity and cosmology},''
  \href{http://dx.doi.org/10.1088/0034-4885/79/10/106901}{{\em Rept. Prog.
  Phys.} {\bf 79} (2016) no.~10, 106901},
\href{http://arxiv.org/abs/1511.07586}{{\tt arXiv:1511.07586 [gr-qc]}}.

\bibitem{Ferraro:2006jd}
R.~Ferraro and F.~Fiorini, ``{Modified teleparallel gravity: Inflation without
  inflaton},'' \href{http://dx.doi.org/10.1103/PhysRevD.75.084031}{{\em Phys.
  Rev.} {\bf D75} (2007)  084031},
\href{http://arxiv.org/abs/gr-qc/0610067}{{\tt arXiv:gr-qc/0610067 [gr-qc]}}.

\bibitem{Bengochea:2008gz}
G.~R. Bengochea and R.~Ferraro, ``{Dark torsion as the cosmic speed-up},''
  \href{http://dx.doi.org/10.1103/PhysRevD.79.124019}{{\em Phys. Rev.} {\bf
  D79} (2009)  124019},
\href{http://arxiv.org/abs/0812.1205}{{\tt arXiv:0812.1205 [astro-ph]}}.

\bibitem{Cai:2011tc}
Y.-F. Cai, S.-H. Chen, J.~B. Dent, S.~Dutta, and E.~N. Saridakis, ``{Matter
  Bounce Cosmology with the f(T) Gravity},''
  \href{http://dx.doi.org/10.1088/0264-9381/28/21/215011}{{\em Class. Quant.
  Grav.} {\bf 28} (2011)  215011},
\href{http://arxiv.org/abs/1104.4349}{{\tt arXiv:1104.4349 [astro-ph.CO]}}.

\bibitem{Bamba:2010wb}
K.~Bamba, C.-Q. Geng, C.-C. Lee, and L.-W. Luo, ``{Equation of state for dark
  energy in $f(T)$ gravity},''
  \href{http://dx.doi.org/10.1088/1475-7516/2011/01/021}{{\em JCAP} {\bf 1101}
  (2011)  021},
\href{http://arxiv.org/abs/1011.0508}{{\tt arXiv:1011.0508 [astro-ph.CO]}}.

\bibitem{Dent:2011zz}
J.~B. Dent, S.~Dutta, and E.~N. Saridakis, ``{f(T) gravity mimicking dynamical
  dark energy. Background and perturbation analysis},''
  \href{http://dx.doi.org/10.1088/1475-7516/2011/01/009}{{\em JCAP} {\bf 1101}
  (2011)  009},
\href{http://arxiv.org/abs/1010.2215}{{\tt arXiv:1010.2215 [astro-ph.CO]}}.

\bibitem{Tamanini:2012hg}
N.~Tamanini and C.~G. Boehmer, ``{Good and bad tetrads in f(T) gravity},''
  \href{http://dx.doi.org/10.1103/PhysRevD.86.044009}{{\em Phys. Rev.} {\bf
  D86} (2012)  044009},
\href{http://arxiv.org/abs/1204.4593}{{\tt arXiv:1204.4593 [gr-qc]}}.

\bibitem{Ferraro:2011ks}
R.~Ferraro and F.~Fiorini, ``{Spherically symmetric static spacetimes in vacuum
  f(T) gravity},'' \href{http://dx.doi.org/10.1103/PhysRevD.84.083518}{{\em
  Phys. Rev.} {\bf D84} (2011)  083518},
\href{http://arxiv.org/abs/1109.4209}{{\tt arXiv:1109.4209 [gr-qc]}}.

\bibitem{Paliathanasis:2014iva}
A.~Paliathanasis, S.~Basilakos, E.~N. Saridakis, S.~Capozziello, K.~Atazadeh,
  F.~Darabi, and M.~Tsamparlis, ``{New Schwarzschild-like solutions in f(T)
  gravity through Noether symmetries},''
  \href{http://dx.doi.org/10.1103/PhysRevD.89.104042}{{\em Phys. Rev.} {\bf
  D89} (2014)  104042},
\href{http://arxiv.org/abs/1402.5935}{{\tt arXiv:1402.5935 [gr-qc]}}.

\bibitem{Daouda:2012nj}
M.~H. Daouda, M.~E. Rodrigues, and M.~J.~S. Houndjo, ``{Anisotropic fluid for a
  set of non-diagonal tetrads in f(T) gravity},''
  \href{http://dx.doi.org/10.1016/j.physletb.2012.07.039}{{\em Phys. Lett.}
  {\bf B715} (2012)  241--245},
\href{http://arxiv.org/abs/1202.1147}{{\tt arXiv:1202.1147 [gr-qc]}}.

\bibitem{Aftergood:2014wla}
J.~Aftergood and A.~DeBenedictis, ``{Matter conditions for regular black holes
  in $f(T)$ gravity},''
  \href{http://dx.doi.org/10.1103/PhysRevD.90.124006}{{\em Phys. Rev.} {\bf
  D90} (2014) no.~12, 124006},
\href{http://arxiv.org/abs/1409.4084}{{\tt arXiv:1409.4084 [gr-qc]}}.

\bibitem{Boehmer:2019uxv}
C.~G. Böhmer and F.~Fiorini, ``{The regular black hole in four dimensional
  Born–Infeld gravity},''
  \href{http://dx.doi.org/10.1088/1361-6382/ab1e8d}{{\em Class. Quant. Grav.}
  {\bf 36} (2019) no.~12, 12LT01},
\href{http://arxiv.org/abs/1901.02965}{{\tt arXiv:1901.02965 [gr-qc]}}.

\bibitem{Bohmer:2011si}
C.~G. Boehmer, T.~Harko, and F.~S.~N. Lobo, ``{Wormhole geometries in modified
  teleparralel gravity and the energy conditions},''
  \href{http://dx.doi.org/10.1103/PhysRevD.85.044033}{{\em Phys. Rev.} {\bf
  D85} (2012)  044033},
\href{http://arxiv.org/abs/1110.5756}{{\tt arXiv:1110.5756 [gr-qc]}}.

\bibitem{Jamil:2012ti}
M.~Jamil, D.~Momeni, and R.~Myrzakulov, ``{Wormholes in a viable f(T)
  gravity},'' \href{http://dx.doi.org/10.1140/epjc/s10052-012-2267-8}{{\em Eur.
  Phys. J.} {\bf C73} (2013)  2267},
\href{http://arxiv.org/abs/1212.6017}{{\tt arXiv:1212.6017 [gr-qc]}}.

\bibitem{Bahamonde:2016jqq}
S.~Bahamonde, U.~Camci, S.~Capozziello, and M.~Jamil, ``{Scalar-Tensor
  Teleparallel Wormholes by Noether Symmetries},''
  \href{http://dx.doi.org/10.1103/PhysRevD.94.084042}{{\em Phys. Rev.} {\bf
  D94} (2016) no.~8, 084042},
\href{http://arxiv.org/abs/1608.03918}{{\tt arXiv:1608.03918 [gr-qc]}}.

\bibitem{Jawad:2015uea}
A.~Jawad and S.~Rani, ``{Lorentz Distributed Noncommutative Wormhole Solutions
  in Extended Teleparallel Gravity},''
  \href{http://dx.doi.org/10.1140/epjc/s10052-015-3386-9}{{\em Eur. Phys. J.}
  {\bf C75} (2015) no.~4, 173},
\href{http://arxiv.org/abs/1504.01657}{{\tt arXiv:1504.01657 [gr-qc]}}.

\bibitem{Horvat:2014xwa}
D.~Horvat, S.~Ilijić, A.~Kirin, and Z.~Narančić, ``{Nonminimally coupled
  scalar field in teleparallel gravity: boson stars},''
  \href{http://dx.doi.org/10.1088/0264-9381/32/3/035023}{{\em Class. Quant.
  Grav.} {\bf 32} (2015) no.~3, 035023},
\href{http://arxiv.org/abs/1407.2067}{{\tt arXiv:1407.2067 [gr-qc]}}.

\bibitem{Boehmer:2011gw}
C.~G. Boehmer, A.~Mussa, and N.~Tamanini, ``{Existence of relativistic stars in
  f(T) gravity},'' \href{http://dx.doi.org/10.1088/0264-9381/28/24/245020}{{\em
  Class. Quant. Grav.} {\bf 28} (2011)  245020},
\href{http://arxiv.org/abs/1107.4455}{{\tt arXiv:1107.4455 [gr-qc]}}.

\bibitem{Ilijic:2018ulf}
S.~Ilijic and M.~Sossich, ``{Compact stars in $f(T)$ extended theory of
  gravity},'' \href{http://dx.doi.org/10.1103/PhysRevD.98.064047}{{\em Phys.
  Rev.} {\bf D98} (2018) no.~6, 064047},
\href{http://arxiv.org/abs/1807.03068}{{\tt arXiv:1807.03068 [gr-qc]}}.

\bibitem{Pace:2017dpu}
M.~Pace and J.~L. Said, ``{A perturbative approach to neutron stars in $f(T,
  \mathcal {T})$ -gravity},''
  \href{http://dx.doi.org/10.1140/epjc/s10052-017-4838-1}{{\em Eur. Phys. J.}
  {\bf C77} (2017) no.~5, 283},
\href{http://arxiv.org/abs/1704.03343}{{\tt arXiv:1704.03343 [gr-qc]}}.

\bibitem{Pace:2017aon}
M.~Pace and J.~L. Said, ``{Quark stars in $f(T, \mathcal {T})$ -gravity},''
  \href{http://dx.doi.org/10.1140/epjc/s10052-017-4637-8}{{\em Eur. Phys. J.}
  {\bf C77} (2017) no.~2, 62},
\href{http://arxiv.org/abs/1701.04761}{{\tt arXiv:1701.04761 [gr-qc]}}.

\bibitem{Ruggiero:2015oka}
M.~L. Ruggiero and N.~Radicella, ``{Weak-Field Spherically Symmetric Solutions
  in $f(T)$ gravity},''
  \href{http://dx.doi.org/10.1103/PhysRevD.91.104014}{{\em Phys. Rev.} {\bf
  D91} (2015)  104014},
\href{http://arxiv.org/abs/1501.02198}{{\tt arXiv:1501.02198 [gr-qc]}}.

\bibitem{DeBenedictis:2016aze}
A.~DeBenedictis and S.~Ilijic, ``{Spherically symmetric vacuum in covariant
  $F(T) = T + \frac{\alpha}{2}T^{2} + \mathcal{O}(T^{\gamma})$ gravity
  theory},'' \href{http://dx.doi.org/10.1103/PhysRevD.94.124025}{{\em Phys.
  Rev.} {\bf D94} (2016) no.~12, 124025},
\href{http://arxiv.org/abs/1609.07465}{{\tt arXiv:1609.07465 [gr-qc]}}.

\bibitem{Krssak:2015oua}
M.~Krššák and E.~N. Saridakis, ``{The covariant formulation of f(T)
  gravity},'' \href{http://dx.doi.org/10.1088/0264-9381/33/11/115009}{{\em
  Class. Quant. Grav.} {\bf 33} (2016) no.~11, 115009},
\href{http://arxiv.org/abs/1510.08432}{{\tt arXiv:1510.08432 [gr-qc]}}.

\bibitem{Golovnev:2017dox}
A.~Golovnev, T.~Koivisto, and M.~Sandstad, ``{On the covariance of teleparallel
  gravity theories},'' \href{http://dx.doi.org/10.1088/1361-6382/aa7830}{{\em
  Class. Quant. Grav.} {\bf 34} (2017) no.~14, 145013},
\href{http://arxiv.org/abs/1701.06271}{{\tt arXiv:1701.06271 [gr-qc]}}.

\bibitem{Hohmann:2017duq}
M.~Hohmann, L.~Järv, M.~Krššák, and C.~Pfeifer, ``{Teleparallel theories of
  gravity as analogue of nonlinear electrodynamics},''
  \href{http://dx.doi.org/10.1103/PhysRevD.97.104042}{{\em Phys. Rev.} {\bf
  D97} (2018) no.~10, 104042},
\href{http://arxiv.org/abs/1711.09930}{{\tt arXiv:1711.09930 [gr-qc]}}.

\bibitem{Hohmann:2018rwf}
M.~Hohmann, L.~Järv, and U.~Ualikhanova, ``{Covariant formulation of
  scalar-torsion gravity},''
  \href{http://dx.doi.org/10.1103/PhysRevD.97.104011}{{\em Phys. Rev.} {\bf
  D97} (2018) no.~10, 104011},
\href{http://arxiv.org/abs/1801.05786}{{\tt arXiv:1801.05786 [gr-qc]}}.

\bibitem{Hohmann:2019nat}
M.~Hohmann, L.~Järv, M.~Krššák, and C.~Pfeifer, ``{Modified teleparallel
  theories of gravity in symmetric spacetimes},''
\href{http://arxiv.org/abs/1901.05472}{{\tt arXiv:1901.05472 [gr-qc]}}.

\bibitem{Golovnev:2018wbh}
A.~Golovnev and T.~Koivisto, ``{Cosmological perturbations in modified
  teleparallel gravity models},''
  \href{http://dx.doi.org/10.1088/1475-7516/2018/11/012}{{\em JCAP} {\bf 1811}
  (2018) no.~11, 012},
\href{http://arxiv.org/abs/1808.05565}{{\tt arXiv:1808.05565 [gr-qc]}}.

\bibitem{Ferraro:2018axk}
R.~Ferraro and M.~J. Guzmán, ``{Quest for the extra degree of freedom in
  $f(T)$ gravity},'' \href{http://dx.doi.org/10.1103/PhysRevD.98.124037}{{\em
  Phys. Rev.} {\bf D98} (2018) no.~12, 124037},
\href{http://arxiv.org/abs/1810.07171}{{\tt arXiv:1810.07171 [gr-qc]}}.

\bibitem{Ferraro:2018tpu}
R.~Ferraro and M.~J. Guzmán, ``{Hamiltonian formalism for f(T) gravity},''
  \href{http://dx.doi.org/10.1103/PhysRevD.97.104028}{{\em Phys. Rev.} {\bf
  D97} (2018) no.~10, 104028},
\href{http://arxiv.org/abs/1802.02130}{{\tt arXiv:1802.02130 [gr-qc]}}.

\bibitem{Farrugia:2018gyz}
G.~Farrugia, J.~L. Said, V.~Gakis, and E.~N. Saridakis, ``{Gravitational Waves
  in Modified Teleparallel Theories},''
  \href{http://dx.doi.org/10.1103/PhysRevD.97.124064}{{\em Phys. Rev.} {\bf
  D97} (2018) no.~12, 124064},
\href{http://arxiv.org/abs/1804.07365}{{\tt arXiv:1804.07365 [gr-qc]}}.

\bibitem{Nunes:2018evm}
R.~C. Nunes, S.~Pan, and E.~N. Saridakis, ``{New observational constraints on
  $f(T)$ gravity through gravitational-wave astronomy},''
  \href{http://dx.doi.org/10.1103/PhysRevD.98.104055}{{\em Phys. Rev.} {\bf
  D98} (2018) no.~10, 104055},
\href{http://arxiv.org/abs/1810.03942}{{\tt arXiv:1810.03942 [gr-qc]}}.

\bibitem{Nunes:2019bjq}
R.~C. Nunes, M.~E.~S. Alves, and J.~C.~N. de~Araujo, ``{Forecast constraints on
  $f(T)$ gravity with gravitational waves from compact binary coalescences},''
\href{http://arxiv.org/abs/1905.03237}{{\tt arXiv:1905.03237 [gr-qc]}}.

\end{thebibliography}\endgroup

\end{document}